\documentclass[journal]{IEEEtran}

\usepackage{graphicx,amssymb,lineno}
\usepackage{amsmath,amsfonts,amssymb}

\usepackage{algorithmic}
\usepackage[usenames]{color}
\usepackage{float}

\usepackage{multirow}

\usepackage[linesnumbered,ruled,vlined]{algorithm2e}

\usepackage{graphicx,graphics,color,epsfig,subfigure,graphpap,rotate}
\usepackage{times, verbatim, subfigure, epsfig, graphicx, latexsym, amsmath}
\usepackage{url}
\usepackage{subfigure}
\begin{document}
%
\title{Device-to-Device Communications Enabled\\
Energy Efficient Multicast Scheduling in
mmWave Small Cells}

\author{Yong~Niu,
        Yu Liu,
        Yong~Li,~\IEEEmembership{Senior Member,~IEEE,}
        Xinlei~Chen,
        Zhangdui Zhong,
        and Zhu~Han,~\IEEEmembership{Fellow,~IEEE}

\thanks{Y. Niu and Z. Zhong are with the State Key Laboratory of Rail Traffic
Control and Safety, Beijing Engineering Research Center of High-speed Railway Broadband Mobile Communications, and the School of Electronic and Information Engineering, Beijing Jiaotong University, Beijing 100044, China (e-mail:
niuy11@163.com).
}

\thanks{Y. Liu, Y. Li, and X.~Chen are with State Key Laboratory on
 Microwave and Digital Communications, Tsinghua National Laboratory for Information
 Science and Technology (TNLIST), Department of Electronic Engineering, Tsinghua
 University, Beijing 100084, China (e-mail: liyong07@tsinghua.edu.cn).}

 \thanks{
Z. Han is with the University of Houston, Houston, TX 77004 USA (e-mail:zhan2@uh.edu), and also with the Department of Computer Science and Engineering, Kyung Hee University, Seoul, South Korea.}


%
\thanks{This study was supported by the Fundamental Research Funds for the Central Universities Grant 2016RC056; and by the State Key Laboratory of Rail Traffic Control and Safety (Contract No. RCS2017ZT009), Beijing Jiaotong University; and by the China Postdoctoral Science Foundation
under Grant 2017M610040; and by the National Natural Science Foundation of China Grants 61725101. The research is partially supported by US NSF CNS-1717454,
CNS-1731424, CNS-1702850, CNS-1646607,ECCS-1547201,CMMI-
1434789,CNS-1443917, and ECCS-1405121.}
}

\maketitle

\begin{abstract}


To keep pace with the rapid growth of mobile traffic demands, dense deployment of small cells in millimeter wave (mmWave) bands has become a promising candidate for next generation wireless communication systems. With a greatly increased data rate from huge bandwidth of mmWave communications, energy consumption should be mitigated for higher energy efficiency. Due to content popularity, many content-based mobile applications can be supported by the multicast service.
mmWave communications exploit directional antennas to overcome high path loss, and concurrent transmissions can be enabled for better multicast service. On the other hand, device-to-device (D2D) communications in physical proximity should be exploited to improve multicast performance.
In this paper, we propose an energy efficient multicast scheduling scheme, referred to as EMS, which utilizes both D2D communications and concurrent transmissions
to achieve high energy efficiency. In EMS, a D2D path planning algorithm establishes multi-hop D2D transmission paths,
and a concurrent scheduling algorithm allocates the links on the D2D paths into different pairings. Then the transmission power of links
is adjusted by the power control algorithm.
Furthermore, we theoretically analyze
the roles of D2D communications and concurrent transmissions in reducing energy consumption.
Extensive simulations under various system parameters demonstrate the superior performance of EMS in terms of energy consumption compared with the state-of-the-art schemes.
Furthermore, we also investigate the choice of the interference threshold to optimize network performance.

\end{abstract}


\section{Introduction}\label{S1}


One of the critical goals for future wireless communication systems is to mitigate the energy consumption in light of a greatly increased data rate. With abundant spectrum in the millimeter wave (mmWave) band, mmWave communications are able to provide multi-gigabit communication services, and thus become a hot topic \cite{elkashlan2015millimeter}. Moreover, research progress on integrated circuits for mmWave communications, including on-chip and in-package antennas, radiofrequency (RF) power amplifiers (PAs), low-noise amplifiers (LNAs), voltage-controlled oscillators (VCOs), mixers, and analog-to-digital converters (ADCs), has paved the way for electronic products
in the mmWave band \cite{rappaport2011state}. There are also several standards defined for indoor wireless personal area networks (WPAN) or
wireless local area networks (WLAN), such as ECMA 387 \cite{ECMA_387}, IEEE 802.15.3c \cite{IEEE_802.15.3c}, IEEE 802.11ad \cite{IEEE_802.11ad}, and IEEE802.11ay \cite{802.11ay}. Densely deploying small cells in the mmWave band underlying the conventional macrocell network has been proposed to
improve network capacity, and this deployment has become a promising candidate for future wireless communication systems. With millions more base stations and billions of connected devices in the 5G era, energy efficiency or energy consumption optimization for mmWave communication systems becomes a critical problem to be investigated.

mmWave communications with higher carrier frequencies experience higher path loss than low carrier frequency communications. For example, the free space path loss at 60 GHz band is 28 decibels (dB) more than that at 2.4 GHz \cite{singh_outdoor}. To combat high channel attenuation, the analog beamforming technique is exploited to synthesize directional antennas in a small platform at both the transmitter and receiver for high antenna gain \cite{beam_training,Beamtraining2,Xiao1}. Consequently, the omnidirectional carrier sensing is disabled, and there is the deafness problem. However, there is less interference between directional links, and concurrent transmissions (spatial reuse) can be exploited to improve network performance in terms of throughput or energy efficiency.


On the other hand, content popularity is found in mobile networks, which follows
the classic Zipf's law \cite{content_popularity}. In other words, most of the requests are for a small amount of content.
Therefore, many content-based applications like TV content streaming, advertising messages broadcasting, and broadcast communication services can be supported by the multicast service, where the base station (BS) provides multiple users (a multicast
group) with the same data \cite{pcds, sdm}. At the same time, there will be many user devices located near to each other in the user-intensive region.
In this case, device-to-device (D2D) communications in physical proximity can be exploited to save power and improve energy efficiency \cite{Yong,Hanbook}. Compared with D2D communications sharing the cellular frequency band, there is less interference between D2D communications in the mmWave band and the cellular systems. In multicast services, the BS needs to serve each user one by one serially in the traditional way, and since the transmission links are adjacent, concurrent transmissions cannot exploited to achieve better network performance. With D2D communications enabled, users with the multicast data are able to forward the multicast data to other users using better D2D channels, and there will be more nonadjacent links, and concurrent transmissions can be enabled to achieve better performance. As in the standard of IEEE 802.11ad \cite{IEEE_802.11ad}, the PBSS (Personal Basic Service Set) is a type of IEEE 802.11 LAN ad hoc network in which stations are able to communicate directly with each other. In the PBSS, one STA is required to assume the role of the PBSS central point (PCP). Thus, we assume the same capability for the BS and user stations in this paper.

In this paper, we study the problem of energy efficient multicast scheduling for mmWave small cells, which exploits concurrent transmissions and
D2D communications to achieve high energy efficiency via power control.
Our contribution is four-fold, and is summarized as follows.

 \begin{itemize}
 \item The optimal multicast scheduling problem with D2D communications and concurrent transmissions considered is formulated into a mixed integer nonlinear program (MINLP), which minimizes the total energy consumption of multicast transmissions by power control with the throughput larger than or equal to that of the serial unicast scheme.

 \item We propose an energy efficient and practical multicast scheduling scheme, called EMS, to solve the formulated problem. EMS consists of D2D path planning algorithm, concurrent scheduling algorithm, and power control algorithm. The D2D path planning algorithm establishes the multi-hop D2D transmission paths. The concurrent scheduling algorithm concurrently schedules the links on the D2D paths into different pairings, while the power control algorithm adjusts the transmission power of links for lower energy consumption with the achieved network throughput ensured.

 \item We demonstrate the roles of D2D communications and concurrent transmissions in reducing energy consumption via theoretical analysis.

 \item Extensive evaluations under various system parameters demonstrate EMS achieves the lowest energy consumption with the throughput ensured compared with other schemes. Based on the results, we also find that D2D communications play a big role in reducing energy consumption in EMS. Moreover, we study the impact of the interference threshold on network performance.

\end{itemize}

We organize the rest of this paper as follows. Section \ref{S2} presents the related work on directional MAC protocols for WPANs or WLANs in the mmWave band. Section \ref{S3} introduces the system model, and analyzes the energy efficient multicast scheduling problem by
an example. In Section \ref{S4}, the optimal energy efficient multicast scheduling problem is formulated into a MINLP. Our proposed EMS scheme is presented in Section \ref{S5}. We theoretically analyze the roles of D2D communications and concurrent transmissions in reducing energy consumption in Section \ref{S5+}.
Performance evaluation of EMS is conducted in Section \ref{S6}. Section \ref{S7} concludes this paper.


\section{Related Work}\label{S2}

There has been some related works on directional MAC protocols for WPANs or WLANs in the mmWave band \cite{Qiao,EX_Region,Qiao_6,Qiao_15,Qiao_7}.
Since the standards of ECMA 387 and IEEE 802.15.3c adopt TDMA, some works are based on TDMA. Cai \emph{et al.} \cite{EX_Region} derived the ER conditions that concurrent transmissions always outperform TDMA for both omni-antenna and directional-antenna models, and proposed the REX scheduling scheme (REX) to achieve significant spatial reuse gain. There are also two protocols based on IEEE 802.15.3c, which exploit concurrent transmissions
to improve performance when the multi-user interference is below a specific threshold. In the scenario of an indoor WPAN, Qiao \emph{et al.} \cite{Qiao} proposed a concurrent transmission scheduling algorithm to maximize the number of flows with the quality of service requirement of each flow satisfied. Furthermore, a multi-hop concurrent transmission scheme is proposed to address the link outage problem and combat huge path loss.
For bursty data traffic, TDMA based protocols may allocate not enough medium time for some flows, while overmuch medium time for others \cite{mao}.


Some centralized scheduling protocols are also proposed for WPANs or WLANs in the mmWave band \cite{mao,Gong,MRDMAC,chenqian,tvt_own,JSAC_own}. Gong \emph{et al.} \cite{Gong} proposed a directional CSMA/CA protocol, which exploits virtual carrier sensing to address the deafness problem. Singh \emph{et al.} \cite{MRDMAC} proposed a multihop relay directional MAC protocol (MRDMAC), which exploits relaying to overcome blockage. The frame based directional MAC protocol (FDMAC) is proposed in \cite{mao}, where the greedy coloring algorithm exploits concurrent transmissions for high efficiency. In the scenario of an IEEE 802.11ad WLAN, Chen \emph{et al.}
\cite{chenqian} proposed a directional cooperative MAC protocol, D-CoopMAC, to coordinate the
uplink channel access. Niu \emph{et
al.} \cite{tvt_own} proposed a blockage robust and efficient directional MAC protocol (BRDMAC) to
overcome the blockage problem by two-hop relaying. In the scenario of heterogeneous cellular networks,
Niu \emph{et al.} \cite{JSAC_own} proposed a
joint transmission scheduling scheme for the radio access and
backhaul of small cells in 60 GHz band (D2DMAC), where a path selection criterion
is designed to enable device-to-device transmissions for performance improvement.

In terms of multicast communication, Naribole \emph{et al.} \cite{sdm}
design, implement, and experimentally evaluate scalable directional multicast (SDM) to train the access point with per-beam per-client RSSI measurements via partially traversing a codebook tree. Based on the training information, a scalable beam grouping algorithm is designed to achieve the minimum multicast group data transmission time. Park \emph{et al.} \cite{IMG} proposed an incremental multicast grouping scheme, which generates
adaptive beamwidths depending on the locations of multicast devices to maximize the sum rate of devices. However, D2D communications are not considered in this scheme. An efficient scheduling scheme for popular content downloading (PCDS) is developed in \cite{pcds}, where users far from the AP obtain the popular content from nearby users via D2D communications. At the same time, concurrent transmissions are also enabled to improve performance.


In terms of energy efficient MAC protocols for wireless networks in the mmWave band, Niu \emph{et al.} \cite{Green} proposed
an energy efficient scheduling scheme for the mmWave backhauling network, which exploits concurrent transmissions to achieve higher energy efficiency. However, D2D communications are not considered in that scheme.


\section{System Overview}\label{S3}

\subsection{System Model}\label{S3-1}

We consider an mmWave small cell with one BS and multiple users (UEs). System time is partitioned into time slots of equal length.
We assume the clocks of UEs are synchronized by the BS, and the BS schedules the medium access of all UEs to accommodate their traffic demands.
In the small cell, directional transmissions are supported via electronically
steerable directional antennas at the BS and UEs. Referring to the personal basic service set (PBSS) in the standard of IEEE 802.11ad,
we assume uniform configuration for the BS and UEs, and one UE is required to assume the role of the BS \cite{IEEE_802.11ad}.
In addition, there is a bootstrapping program run in the system, from which the BS knows the up-to-date network topology and the locations of UEs \cite{bootstrapping,location_1}.
The network topology can be obtained by the neighbor discovery schemes in \cite{bootstrapping,neighbor discovery,neighbor discovery 2,neighbor discovery 3,neighbor discovery 4}. Location information of nodes can be obtained based on wireless channel signatures, such as angle of arrival (AoA), time difference of arrival (TDoA), or the received signal strength (RSS) \cite{location_1, location_2, location_3, location_4, location_5}. In our system, the bootstrapping program adopts the direct discovery scheme to discover the network topology \cite{bootstrapping}. In the direct discovery scheme, a node is in the transmitting or receiving state at the beginning of each time slot. In the transmitting state, a node transmits a broadcast packet with its identity in a randomly chosen direction. In the receiving state, a node listens for broadcast packets from a randomly chosen direction. If a collision happens, the node fails to discover any neighbor; otherwise, if the transmitter is unknown, the receiver discovers a new neighbor by recording the angle of arrival (AOA) and the transmitter's identity. After the direct discovery, the nodes report discovered neighbors to the BSs. At the same time, the BSs obtain the location information of UEs by a maximum-likelihood (ML) classifier based on changes in the second-order statistics and sparsity patterns of the beamspace multiple input multiple output (MIMO) channel matrix \cite{location_1}.

On the other hand, for mmWave communication systems, it has been proposed to separate the C-plane (control plane) and the U-plane (user plane), where mmWave communications are used for data transmissions of users, and control signalling information is transmitted using high-quality lower frequency bands to handle mobility \cite{C-U,D2D-Qiao}. In this case, advanced localization techniques in lower frequency bands can be utilized to obtain the location of nodes \cite{MoeWin-location1,MoeWin-location}. As shown in \cite{MoeWin-location1}, through spatio-temporal cooperation, high-accuracy location information can be obtained. With the location information of nodes, the mmWave beam alignment overhead can be significantly reduced \cite{mmW-V2V}, and we assume after the bootstrapping program, the BS and UEs are able to point their beams towards each other.

In our system, the multicast service is completed in a frame \cite{mao}, and each frame consists of the scheduling part and the transmission part.
Considering the relative low mobility of UEs, the bootstrapping program will be executed periodically, and the network status will be updated periodically. If the network status changes during a frame, there will be failed transmissions. In the next frame, the receivers of failed transmissions will report the failed transmissions to the BS, and the network status will be updated before next frame. Considering the receivers may be lost from the BS due to blockage by obstacles, the receivers may use the low frequency band to report the network status changes to the BS.
With a transmission rate at the order of gigabit per second, the information exchange and training between the BS and UEs or two UEs can be completed in a short time, and the transmission part occupies the most of a frame. Thus, in this paper, we focus on the energy consumption in the transmission part. We assume half-duplex nodes for the BS and UEs, and
at most one connection can be supported simultaneously for each node.




As shown in \cite{NLOS,NLOS2,NLOS3}, non-line-of-sight (NLOS) transmissions suffer higher propagation loss than line-of-sight
(LOS) transmissions for mmWave small cells. Due to the LOS path is strongest, LOS transmissions can also improve the power efficiency.
Suffering from a shortage of multipath for NLOS transmissions, we only consider the LOS transmissions in this paper, and the directional LOS link budget is calculated
according to the additive white Gaussian noise channel model \cite{MRDMAC}.

The directional link from nodes $i$ to $j$ is denoted by $(i,j)$. After the beamforming process, nodes $i$ and $j$ point their antennas towards each other. Then received power ${{P^r_{ij}}}$ (mW) at node $j$ from node $i$ can be estimated according to the path loss model, which can be expressed as
\begin{equation}
{P^r_{ij}} = {k_0}{G_t}(i,j){G_r}(i,j){l_{ij}^{ - \tau }}{P_t},
\end{equation}
where we denote the transmission power by ${{P_t}}$ (mW), the distance between nodes $i$ and $j$ by ${{l_{ij}}}$ (m), the path loss exponent by $\tau $, the transmit antenna gain of node $i$ in the direction of $i \to j$ by ${G_t}(i,j)$, the receive antenna gain of node $j$ in the direction of $i \to j$ by ${G_r}(i,j)$, and $k_0$ is a constant coefficient and $k_0 \propto {(\frac{\lambda_c }{{4\pi }})^2}$ ($\lambda_c $ is the wavelength) \cite{Qiao}.

On the other hand, there is less interference between directional links. In this case, we can exploit concurrent transmissions to improve network performance. Adopting the interference model in \cite{Qiao}, for links $(u,v)$ and $(i,j)$,
we can obtain the received interference power at node $j$ from node $u$ as
\begin{equation}
I_{uvij} =\rho k_0{G_t}(u,j){G_r}(u,j){l_{uj}}^{ - \tau }{P_t},
\end{equation}
where $\rho$ is related to the cross correlation of signals from different links \cite{Qiao}.
Then we can obtain the interference power ${{I_{ij}}}$ as
\begin{equation}
{I_{ij}} = \sum\limits_{(u,v) \in {\mathbb{C}_{ij}}} {{I_{uvij}}},
\end{equation}
where we denote the set of links that transmit concurrently with link $(i,j)$ by $\mathbb{C}_{ij}$.

At the same time, adjacent links cannot be scheduled for concurrent transmissions due to the half-duplex assumption.
Therefore, links that are adjacent to link $(i,j)$ are not included in $\mathbb{C}_{ij}$. Then the received signal to interference plus noise ratio (SINR) at receiver $j$ can be expressed as
\begin{equation}
{\Gamma_{ij}} = \frac{{{k_0}{G_t}(i,j){G_r}(i,j){l_{ij}}^{ - \tau }P_t}}{{{N_0}W + \rho \sum\limits_{(u,v) \in {\mathbb{C}_{ij}}} {k_0{G_t}(u,j){G_r}(u,j){l_{uj}}^{ - \tau }{P_t}} }},
\end{equation}
where we denote the bandwidth by $W$ (Hz), and the one-sided
power spectra density of white Gaussian noise by ${{N_0}}$ (mW/Hz) \cite{Qiao}.
Considering the reduction of multipath effect for directional mmWave links \cite{MRDMAC},
link $(i,j)$ is able to achieve a data rate of
\begin{equation}
\begin{array}{l}
{R_{ij}} = \eta W{\log _2}\left(1 + \frac{{{k_0}{G_t}(i,j){G_r}(i,j){l_{ij}}^{ - \tau }P_t}}{{{N_0}W + \rho \hspace{-0.4cm}\sum\limits_{(u,v) \in {\mathbb{C}_{ij}}}\hspace{-0.4cm} {k_0{G_t}(u,j){G_r}(u,j){l_{uj}}^{ - \tau }{P_t}} }}\right),
\end{array}\label{rate}
\end{equation}
where $\eta  \in (0,1)$ denotes the transceiver design efficiency \cite{Qiao}.

In the system, we consider the BS transmits the multicast traffic
to a multicast group. EMS exploits both
concurrent transmissions and D2D transmissions in close proximity to reduce energy consumption while ensuring the network throughput.
We illustrate the operation of EMS in a small cell of five users in Fig. \ref{fig:EMS operation} (b). EMS is frame based, and each frame has two parts, scheduling part and transmission part \cite{mao}. In the first stage of scheduling part, BS obtains the multicast traffic and corresponding multicast group from upper layers, which takes time ${t_{m}}$; Then in the second stage, the BS computes a schedule to accommodate the multicast
traffic demands, which takes time ${t_{sch}}$; in the third stage, the BS pushes the schedule to the users in the multicast group, which takes time ${t_{push}}$. In the second stage of the scheduling part, after establishing the D2D paths, beam training between D2D pairs on the paths is executed
to select beams for the transmitters and receivers.
In the transmission part, all nodes transmit according to the schedule until the multicast traffic demands are accommodated. We define the period during which a group of concurrent links are activated as one pairing, and multiple pairings may exist in the transmission part. In the following section, we discuss the key mechanisms in the problem by this example.

\subsection{Problem Overview}\label{S3-2}

\begin{figure}[htbp]
\begin{minipage}[t]{1\linewidth}
\centering
\includegraphics[width=7.6cm]{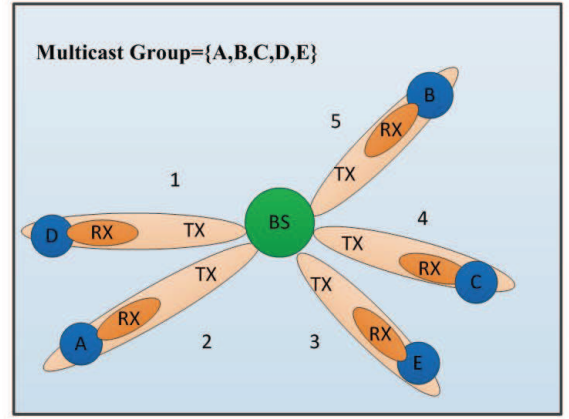}
\centerline{\small (a) Network Topology and the serial unicast scheme operation}
\end{minipage}%
\vfill\vspace{0.6cm}
\begin{minipage}[t]{1\linewidth}
\centering
\includegraphics[width=8.8cm]{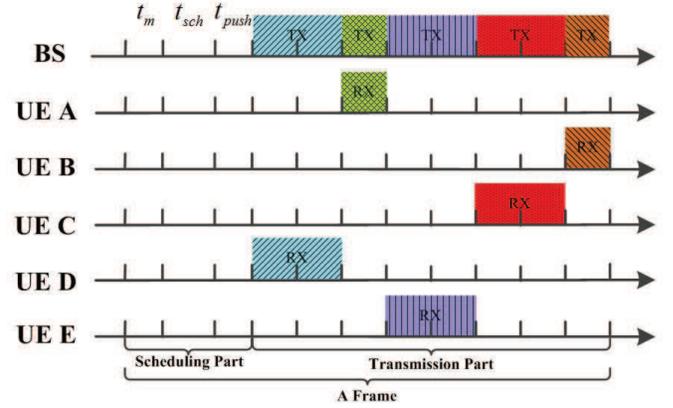}
\centerline{\small (b) Timeline operation of the serial unicast scheme}
\end{minipage}
\caption{An example of the serial unicast scheme operation in a small cell of five users.}
\label{fig:serial operation} 
\vspace*{-3mm}
\end{figure}

\begin{figure}[htbp]
\begin{minipage}[t]{1\linewidth}
\centering
\includegraphics[width=7.6cm]{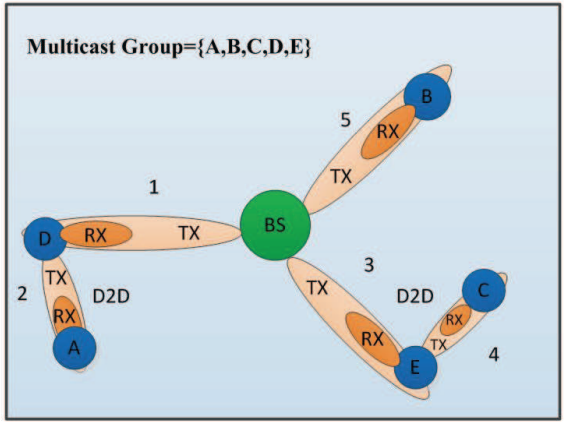}
\centerline{\small (a) Network Topology and EMS operation}
\end{minipage}%
\vfill\vspace{0.6cm}
\begin{minipage}[t]{1\linewidth}
\centering
\includegraphics[width=8.8cm]{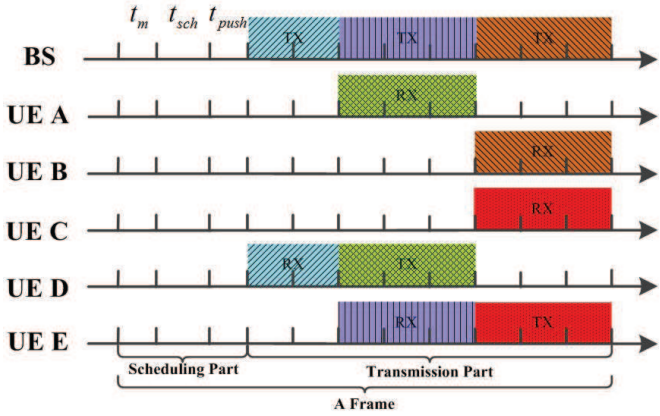}
\centerline{\small (b) Timeline operation of EMS}
\end{minipage}
\caption{An example of EMS operation in a small cell of five users.}
\label{fig:EMS operation} 
\vspace*{-3mm}
\end{figure}

In Fig. \ref{fig:serial operation}, we show the multicast service scenario by the serial unicast scheme in a small cell of five users.
The multicast group is UEs A, B, C, D, and E. In the serial unicast scheme, the BS serially directs its directional beam towards each user, and the multicast data is transmitted to each user, as shown in Fig. \ref{fig:serial operation} (a). In Fig. \ref{fig:serial operation} (b), we illustrate the timeline operation of the serial unicast scheme.

As comparison, we plot the multicast service by EMS in Fig. \ref{fig:EMS operation}. In the EMS scheme, we select three transmission paths, BS $\to$ B, BS $\to$ D $\to$ A, and BS $\to$ E $\to$ C as in Fig. \ref{fig:EMS operation} (a). Fig. \ref{fig:EMS operation} (b) gives the timeline operation of the schedule, which has three pairings in the transmission part. In the first pairing, the BS transmits the multicast data to D. In the second pairing, the BS transmits the multicast data to E, and D transmits to A. In the third pairing, link BS $\to$ B and link E $\to$ C are activated to distribute the multicast data to B and C. We can observe that in the second and third pairing, two links transmit concurrently. At the same time, the two-hop D2D transmission paths, BS $\to$ D $\to$ A and BS $\to$ E $\to$ C, are established.

As we can see, the schedule completes the transmission in the same total time as the serial unicast scheme, which indicates that the achieved throughput of EMS is not less than that by the serial unicast scheme.
However, each link in the schedule gets more or equal time slots for transmission than those in the serial unicast scheme. With more time slots for each transmission, lower transmission power can be achieved while completing the transmission of multicast data, and the energy consumption can be reduced accordingly. With the same number of time slots allocated to the multicast service in EMS, each transmission can obtain the most time slots, and the transmission power can be reduced as much as possible. For example, if the transmission time doubles compared with the serial unicast scheme, the transmission power usually can be reduced by more than half. Consequently, the energy consumption can be reduced. From the example, we can observe that there are two key mechanisms to be exploited to reduce energy consumption. The first one is D2D communications, and multi-hop D2D transmission paths should be established. The second one is concurrent transmissions, and interference between concurrent links should be managed appropriately to fully reap the benefits of concurrent transmissions.



\section{Problem Formulation And Analysis}\label{S4}

To achieve as high energy efficiency as possible, we minimize the energy consumption of multicast transmission with the throughput ensured. With the advantages of D2D communications and concurrent transmissions fully exploited, the transmit power of each transmission can be adjusted to achieve lower energy consumption. Now, we formulate the problem of optimal multicast scheduling in terms of energy consumption with D2D communications and concurrent transmissions enabled.

\subsection{Problem Formulation} \label{S4-1}

We consider the multicast traffic in a mmWave small cell. There is one multicast group in the network, and we denote the traffic demand for the multicast group by $D$. $\mathbb{U}$ denotes the set of users in the multicast group. For each user $u \in \mathbb{U}$, we denote the transmit node that serves $u$ by $s_u$. Since D2D communications are enabled, $s_u$ may be the BS or other users. We assume the schedule for the multicast transmission period has $K$ pairings, and the number of time slots for the $k$th pairing is denoted by $\delta ^k$ \cite{mao}. The duration of one time slot is denoted by $\Delta$. The binary variable $a_u^k$ is defined to indicate whether the multicast transmission for user $u$ is scheduled in the $k$th pairing. If it is, $a_u^k$ is equal to 1; otherwise, $a_u^k$ is equal to 0. We denote the transmission power of $s_u$ to $u$ by $P_t^u$. Then from (\ref{rate}), the achievable transmission rate for user $u$ in the $k$th pairing can be calculated as


\begin{equation}
\begin{array}{l}
{R_{{s_u}u}^k} \hspace{-0.1cm}= \hspace{-0.1cm}\eta W{\log _2}\left(1\hspace{-0.1cm} +\hspace{-0.1cm} \frac{{a_u^k{k_0}{G_t}({s_u},u){G_r}({s_u},u){l_{{s_u}u}}^{ - \tau }P_t^u}}{{{N_0}W + \rho\hspace{-0.6cm} \sum\limits_{(s_v,v) \in {\mathbb{C}_{{s_u}u}}} \hspace{-0.6cm} {a_v^kk_0{G_t}(s_v,u){G_r}(s_v,u){l_{s_vu}}^{ - \tau }{P_t^v}} }}\right),
\end{array}\label{}
\end{equation}
where we use $(s_v,v)$ to denote another link scheduled concurrently with link $(s_u,u)$ in the $k$th pairing.

To achieve high energy efficiency, we try to minimize the energy consumption with the achieved throughput not less than that of the serial unicast scheme. In the serial unicast scheme, the BS transmits the multicast data to each user in the multicast group serially. Thus, the objective function is expressed as
\begin{equation}
\sum\limits_{u = 1}^{|\mathbb{U}|} {\sum\limits_{k = 1}^K {P_t^u  a_u^k  \delta ^k  \Delta} }. \label{OBJ}
\end{equation}
We can observe that the objective function is defined as the total energy consumed to accommodate all users in the multicast group. In our scheme, the transmitter may be the BS or other users via D2D communications.

The constraints of this energy consumption minimization problem is analyzed as follows. First, to reduce complexity and beamforming overhead, we consider the case where the transmission for each user is scheduled only once in the transmission period, which is expressed as

\begin{equation}
\sum\limits_{k = 1}^K {a_u^k}  = 1, \ \ \forall\;  u.  \label{cons1}
\end{equation}

Second, due to the half-duplex assumption, adjacent links cannot be
scheduled concurrently \cite{mao}, which can be expressed as

\begin{equation}
a_u^k + a_v^k \le 1, \  {\rm{if \;links}}\; (s_u,u)\; {\rm{and}} \;(s_v,v)\; {\rm{are\; adjacent}};\label{cons2}
\end{equation}

Third, the multicast demand of each user should be accommodated by the schedule, which can expressed as

\begin{equation}
\sum\limits_{k = 1}^K{R_{{s_u}u}^k  \delta ^k  \Delta }  \ge D,\ \forall\;  u.\label{cons3}
\end{equation}

Fourth, to exploit D2D communications to improve energy efficiency, only the user with the multicast data is able to serve other users. Thus, the multicast transmission for $s_u$ should be scheduled prior to the multicast transmission for user $u$, i.e.,

\begin{equation}
\sum\limits_{k = 1}^{{K^*}}{a_{s_u}^k} \ge \sum\limits_{k = 1}^{{K^*}} {a_{u}^k},\ \forall\;  u,\ {K^*} = 1 \sim K. \label{cons4}
\end{equation}

Fifth, the transmission power of $s_u$ for user $u$ should not exceed the maximum allowed transmission power, denoted by $P_{max}$, which is expressed as follows

\begin{equation}
P_t^u  \le P_{max}, \ \forall\;  u. \label{cons5}
\end{equation}

Finally, to achieve high energy efficiency with the throughput ensured, we require the achieved throughput should be greater than or equal to that achieved by the serial unicast scheme. The number of time slots needed for the serial unicast scheme to complete the multicast service is denoted by $T_s$. With the multicast demand fixed as $D$, we can infer that this throughput requirement is equivalent to the constraint in terms of the occupied number of time slots below. It can be expressed as

\begin{equation}
\sum\limits_{k = 1}^K {{\delta ^k}} \le T_s. \label{cons6}
\end{equation}

Therefore, the problem of optimal multicast scheduling (P1) can be formulated as

\begin{equation}\hspace{-2.8cm}
({\rm{P1}})\ \ \min \sum\limits_{u = 1}^{|\mathbb{U}|} {\sum\limits_{k = 1}^K {P_t^u  a_u^k  \delta ^k  \Delta} },\label{obj1}
\end{equation}

\hspace{1.50cm}s. t.
\hspace{0.2cm}Constraints (\ref{cons1})--(\ref{cons6}).

In problem P1, we can observe that the objective function (\ref{OBJ}) and constraint (\ref{cons3}) have nonlinear terms, especially ${R_{{s_u}u}^k}$ in constraint (\ref{cons3}) has a complex form. Considering the binary variable $a_{u}^k$, the integer variable $\delta^k$, and the real variable $P_t^u$, problem (P1) is a mixed integer nonlinear program (MINLP) and is more complex than the NP-complete 0--1 Knapsack problem \cite{Knapsack}.
Considering the complex form in the objective function (\ref{OBJ}) and constraint (\ref{cons3}), relaxation techniques like the Reformulation-Linearization Technique (RLT) cannot be applied to the nonlinear terms, and it is even difficult to obtain approximate solutions of the original problem \cite{mao}. To achieve a practical solution, we propose the energy efficient multicast scheduling scheme for problem P1 in the following.



\section{Energy Efficient Multicast Scheduling Scheme}\label{S5}


In this section, we propose the energy efficient multicast scheduling scheme, EMS, for the formulated problem. Both D2D communications and concurrent transmissions are enabled in EMS to improve the energy efficiency. First, we propose a D2D path planning algorithm to establish the multi-hop D2D transmission paths. Then a concurrent scheduling algorithm is proposed to schedule the links on the D2D paths concurrently into each pairing with the interference controlled. Finally, a power control algorithm adjusts the transmission power to realize energy consumption reduction.

\subsection{D2D Path Planning Algorithm}\label{S5-1}


The advantage of D2D communications relies on the better channel conditions between devices in physical proximity. D2D communications between users nearby are preferred due to less propagation loss for saving energy. The D2D path planning algorithm establishes multiple D2D transmission paths from the BS, and by finding the nearest user to the last user on one of the allocated D2D transmission paths, this path is extended by including this new user. If one unallocated user is nearest to the BS, one new path from the BS to this user will be established. The number of hops on each path cannot exceed a predetermined value.

We denote the BS by $A$, and the set of selected D2D paths by $\mathbb{P}$. The maximum hop number for each path $p\in\mathbb{P}$ is denoted by $H_m$. For each path $p\in \mathbb{P}$, the last node on $p$ is denoted by $L_p$. The set of last nodes of paths in $\mathbb{P}$ is denoted by $\mathbb{P}_L$. In our algorithm, $\mathbb{P}_L$ represents the set of nodes with the multicast traffic and the ability to serve other users by D2D communications. Since the AP $A$ is the source with the multicast traffic, $A$ is included in $\mathbb{P}_L$.

\begin{algorithm}[htbp]
 \DontPrintSemicolon
 \caption{D2D Path Planning Algorithm.}\label{alg: multicast-D2D-E}
  \textbf{Input:} the multicast group $\mathbb{U}$; \\

  \textbf{Initialization:}    $\mathbb{P}_L=\{A\}$; $\mathbb{P}=\emptyset$;\\
    \While {$|\mathbb{U}| > 0$}
    {

        \For {{\rm{each node}} $i\in \mathbb{P}_L$  }
        {
        Find user $u$ with the shortest distance to node $i$.\\
        $r_i=l_{ui}$.\\
        $c_i=u$.\\
        }
        Find node $s\in \mathbb{P}_L$ with the minimum $r_s$;\\

        \If{$s==A$}
        {
          $\mathbb{P}=\mathbb{P}\cup\{A\to c_s\}$;

        }
        \Else
        {Find the path $p\in \mathbb{P}$ with $L_p=s$;\\

        Update $p$ by extending $p$ to $c_s$;\\

        $\mathbb{P}_L=\mathbb{P}_L-s$;\\
        }
        $\mathbb{U}=\mathbb{U}- c_s$;\\
        \If{$H(p)<H_m$}
        {$\mathbb{P}_L=\mathbb{P}_L\cup c_s$;\\}

    }

$\mathbf{Return}\ \ \mathbb{P}$.
\end{algorithm}


The pseudo-code of the D2D path planning algorithm is presented in Algorithm \ref{alg: multicast-D2D-E}. The algorithm iteratively allocates each UE on the D2D transmission paths until all UEs are allocated, as in line 3. The transmission paths in $\mathbb{P}$ extend to unallocated UEs by searching the nearest neighbors of their last nodes, as in lines 4--7.
Line 8 obtains node $s$ and its UE with the shortest distance. When node $s$ is the BS, a new path from the BS to the selected UE is generated in $\mathbb{P}$, as in lines 9--10. When node $s$ is not the BS, the algorithm extends the transmission paths in $\mathbb{P}$ to the nearest unallocated UE from the last node, and removes node $s$ from $\mathbb{P}_L$, as in lines 11--14. In line 15, the selected UE is removed from the $\mathbb{U}$. Thus, the selected UE will not be on another D2D path, and will be only scheduled once, which is required by constraint (\ref{cons1}).
In lines 16--17, the selected UE is added to $\mathbb{P}_L$ if the hop number of the path is less than $H_m$.



For the example in Fig. \ref{fig:EMS operation}, the D2D path planning algorithm establishes three transmission paths, BS $\to$ D $\to$ A, BS $\to$ B, and BS $\to$ E $\to$ C, for the multicast group of $\{A, B, C, D, E\}$.
From Algorithm \ref{alg: multicast-D2D-E}, we can observe that the outer while loop has $|\mathbb{U}|$ iterations, and the inner for loop also has at most $|\mathbb{U}|$ iterations. Thus, the worst case computational complexity of the D2D path planning algorithm is $\mathcal{O}(|\mathbb{U}|^2)$, which can be implemented in practice.

\subsection{Concurrent Multi-Hop Scheduling}\label{S5-2}

After obtaining the D2D transmission paths by Algorithm \ref{alg: multicast-D2D-E}, the advantage of concurrent transmissions should be further exploited to improve energy efficiency. Thus the concurrent multi-hop scheduling algorithm is proposed to schedule the links on the D2D transmission paths into the transmission period. The algorithm controls the interference via the contention graph, and the maximum independent set (MIS)
is utilized to achieve high efficiency.


\subsubsection{Contention Graph}\label{S5-2-a}

As in \cite{Green,yun_1}, we adopt the contention graph to model the contention relationship among links. Each vertex in the contention graph represents one link in the network. There will be contention (one edge) between two vertices if severe interference between these two links exists.
In other words, if the interference between two links is less than a predetermined threshold, we assume no contention between these two links, and concurrent transmissions of these two links are allowed. Instead, if there is severe interference between links, their concurrent transmissions are disabled.

Concretely, we construct the contention graph as follows. To simplify denotation, we denote link $(s_u,u)$ as vertex $u$.
For links $(s_u,u)$ and $(s_v,v)$, we define the maximum of the interference between them as the weight of the edge between them, i.e.,
\begin{equation}
{W_{uv}} = \max \{  {P^r_{s_{u}v}}, {P^r_{s_{v}u}} \}.
\end{equation}
The transmission power here is the maximum transmission power $P_{max}$. Then the interference threshold $\sigma$ is defined to control the interference. If ${{{W_{uv}}}/{{{P_{max}}}}} < \sigma $ for links $(s_u,u)$ and $(s_v,v)$, no edge exists between these two vertices.
Otherwise, there will be contention between these two links. Since concurrent transmissions for adjacent links are disabled, there will always be contention between adjacent links, which is required by constraint (\ref{cons2}).

\subsubsection{MIS based Multihop Scheduling Algorithm}\label{S5-2-b}


With the contention graph constructed, a maximum independent set (MIS) based multihop scheduling algorithm is proposed to allocate the links on the D2D paths to different pairings. To fully reap the benefits of concurrent transmissions, the MIS based multihop scheduling algorithm schedules as many links into each pairing as possible by obtaining the maximum independent set of the contention graph. With more concurrent transmissions enabled, more time slots can be allocated to each pairing, and lower energy consumption can be achieved by power control.
The MIS of the contention graph is a set of isolated vertexes (links) with the maximum cardinality \cite{yun_1}. Since it is NP-complete to obtain the MIS of a general graph, the minimum-degree greedy algorithm is exploited to approximate the maximum independent set. In
\cite{yun_1_15}, a performance ratio of $(\Omega  + 2)/3$ can be achieved by the minimum-degree greedy algorithm
for the graphs of degree bounded by $\Omega $. On the other hand, as required by constraint (\ref{cons4}), only the first unscheduled links on the D2D paths can be scheduled in the current pairing.

The contention graph constructed by these links (vertices) in the $t$th pairing is denoted by $G_t(V_t,E_t)$, where $V_t$ denotes the set of vertices, and $E_t$ denotes the set of edges. Two vertices are referred to as neighbors if one edge exists between them. We denote the set of neighboring vertices for any vertex $v \in V_t$ by $N(v)$. The degree for any vertex $v \in V_t$ is denoted by $d(v)$. We denote the set of links on paths in $\mathbb{P}$ by $V$. The set of links scheduled in the $t$th pairing is denoted by $V^t$, and the set of unvisited and candidate links for the $t$th pairing is denoted by $V_u^t$.

As the scheduling process goes on, the algorithm iteratively finds the maximum independent set from $G_t(V_t,E_t)$ for each pairing, and $G_t(V_t,E_t)$ is constructed from the first unscheduled links on the D2D paths from $\mathbb{P}$ in the
$t$th pairing. Then the links in the maximum independent set are scheduled in the same pairing for concurrent transmissions. After the scheduling of each pairing, the contention graph is updated for next pairing after removing the scheduled links from $V$ and $\mathbb{P}$.

Algorithm \ref{alg:MIS-h} presents the pseudo-code of the MIS based multihop scheduling algorithm.
The algorithm iteratively schedules the links in $V$ into each pairing, as shown in line 4. In the scheduling for each pairing, the algorithm obtains the first unscheduled links on paths of $\mathbb{P}$, and construct the contention graph for this pairing, as indicated by line 7. In this way, preceding links on each path will be scheduled prior to links behind, which is required by constraint (\ref{cons4}).
Then from line 8 to line 12, the minimum-degree greedy algorithm obtains the MIS for each pairing. In lines 13--14, the set of scheduled links $V^t$ is subtracted from $V$ and $\mathbb{P}$. Since $V$ and $\mathbb{P}$ are updated in lines 13 and 14, the set of first-hop links on the D2D paths is the set of first unscheduled links on the D2D paths, as indicated in line 7.


\begin{algorithm}[htbp]
 \DontPrintSemicolon
 \caption{MIS Based Multihop Scheduling.}\label{alg:MIS-h}
  \textbf{Input:} The set of selected D2D paths, $\mathbb{P}$; \\
                  The set of links on paths in $\mathbb{P}$, $V$;\\
  \textbf{Initialization:}  $t$=0;\\
    \While {$|V| > 0$}
    {
        $t$=$t$+1;  \\
        Set ${V^t} = \emptyset $; \\
        Obtain the set of first-hop links on the D2D paths from $\mathbb{P}$ and $V$, $V_t$;\\
        Set $V_u^t$ with $V_u^t=V_t$; \\
        \While {$|{V_u^t}|>0$ }
        {
            Obtain $v\in V_u^t$ such that $d(v) = \mathop {\min }\limits_{w \in V_u^t} d(w)$; \\
            $V^t=V^t \cup v$;\\
            $V_u^t = V_u^t - \{{v}\cup N(v)\}$;\\
        }
    $V=V-V^t$;\\
    Remove the links in $V^t$ from $\mathbb{P}$;\\
    }

$\mathbf{Return}\ V^t$ of each pairing.
\end{algorithm}


For the example in Fig. \ref{fig:EMS operation}, the MIS based multihop scheduling algorithm obtains three pairings. In the first pairing, there is only one link, BS $\to$ D.
In the second pairing, there are two links, BS $\to$ E and D $\to$ A. In the third pairing, there are two links, BS $\to$ B and E $\to$ C.
From Algorithm \ref{alg:MIS-h}, we can observe that the outer while loop has $|V|$ iterations, and the inner while loop also has at most $|V|$ iterations. Thus, the computational complexity of the MIS based multihop scheduling algorithm is $\mathcal{O}(|V|^2)$, which can also be implemented in practice.

\subsection{Power Control Algorithm}\label{S5-3}

With links on the paths scheduled into each pairing by Algorithm
\ref{alg:MIS-h}, a power control algorithm is proposed to adjust the transmission power of links for lower energy consumption, which is also used in \cite{Green}. To ensure the throughput achieved by our scheme not less than the serial unicast scheme, we require the number of time slots occupied by our scheme not larger than that by the serial unicast scheme, $T_s$. With D2D communications and concurrent transmissions enabled by our scheme, the number of time slots allocated to each link by our scheme is larger than that by the serial unicast scheme with the same number of total occupied time slots.
With more time slots occupied by each link, lower transmission power can be achieved, and thus the energy consumption can be reduced.
For example, if we have twice as much time for transmission, half transmission rate is needed to ensure the throughput.
Then from the Shannon's channel capacity, the transmission power $P_t$ is proportional to $({2^{R/W}} - 1)$. Under relatively high SINR and low interference, $P_t$ can be reduced by more than half. On the other hand, the better channel conditions provided by D2D links can also help to achieve lower energy consumption. Consequently, the energy efficiency can be improved due to reduced energy consumption and ensured throughput.

In Algorithm \ref{alg:PC-D2D}, we present the pseudo-code of the power control algorithm.
For simplicity, we use $u\in V^k$ to denote the link $(s_u,u)$ in the $k$th pairing. When links' transmission power is equal to $P_{max}$,
the transmission rate of link $(s_u,u)$ can be obtained as
\begin{equation}
\begin{array}{l}
R_{{s_u}u}' \hspace{-0.1cm}=\hspace{-0.1cm} \eta W{\log _2}\hspace{-0.1cm}\left(1\hspace{-0.1cm} +\hspace{-0.1cm} \frac{{ {k_0}{G_t}({s_u},u){G_r}({s_u},u){l_{{s_u}u}}^{ - \tau }P_{max}}}{{{N_0}W + \rho \hspace{-0.4cm}\sum\limits_{v \in {V^k{\backslash \{ u\}}}}\hspace{-0.4cm} { k_0{G_t}(s_v,u){G_r}(s_v,u){l_{s_vu}}^{ - \tau }{P_{max}}} }}\right).
\end{array}\label{}
\end{equation}
The number of time slots needed for it to complete the multicast transmission as required by constraint (\ref{cons3}), $T^k_u$, can be calculated as

\begin{equation}
T^k_u=     \frac{{D}}{{R_{{s_u}u}' \Delta}}.
\end{equation}
Thus, we can obtain the maximum needed number of time slots for links in the $k$th pairing as ${T^k} = \mathop {\max \{ }\limits_{u \in {V^k}} T_u^k\}$. To reduce the energy consumption as much as possible, we should exploit the time slots available fully while ensuring the throughput as indicated by constraint (\ref{cons6}).
Thus the $T_s$ time slots is distributed proportionally to each pairing according to $T^k$. Thus the number of time slots for the $k$th pairing $\delta^k$ can be expressed as
\begin{equation}
{\delta ^k} = \left\lfloor {\frac{{{T^k}}}{{\sum\limits_k {{T^k}} }} \cdot T_s} \right\rfloor,\label{condition}
\end{equation}
where the floor operation is on pairings before the final pairing, and the remaining time slots are allocated to the final pairing.

\begin{algorithm}[htbp]
 \DontPrintSemicolon
 \caption{Power Control Algorithm.}\label{alg:PC-D2D}
  \textbf{Input:} The set of links scheduled in each pairing, $V^k$; \\
        \hspace{0.95cm} The number of pairings, $K$;\\
  \textbf{Initialization:}  $k$=0;\\
    \While {$k<K$}
    {
        $k$=$k$+1;  \\
        \For {{\rm{each link}} $u \in V^k$}
        {
            Calculate its transmission rate under $P_{max}$, $R_{{s_u}u}'$; \\
            Obtain the number of time slots to complete multicast transmission by $R_{{s_u}u}'$, $T_u^k$;\\
        }
    Obtain ${T^k} = \mathop {\max \{ }\limits_{u \in {V^k}} T_u^k\}$;\\
    }
   \hspace{0.4cm} $k$=0;\\
    \While {$k<K$}
    {
        $k$=$k$+1;  \\
        Calculate the number of time slots for the $k$th pairing, $\delta ^k$;\\
        \For {{\rm{each link}} $u \in V^k$}
        {
            Calculate the transmission rate to complete multicast transmission, $R_{{s_u}u}''$;\\
            Calculate the transmission power, $P_t^u$;\\
        }
    }

$\mathbf{Return}\ \delta ^k$ of each pairing and $P_t^u$ of each link.
\end{algorithm}

After allocating time slots for each pairing, we can activate the transmission of each link during the whole period of its corresponding pairing to reduce transmission power. The number of time slots allocated to link $(s_u,u)$ in the serial unicast scheme is denoted by $\theta_u$.
Since the threshold $\sigma$ keeps the interference between concurrent links low, ${\delta ^k} \ge {T^k} \ge T_u^k \ge \theta_u$ holds generally, and more time slots can be allocated to each link compared with the serial unicast scheme.
For link $(s_u,u)$ scheduled in the $k$th pairing, its needed transmission rate to complete multicast transmission can be expressed as
\begin{equation}
R_{{s_u}u}'' = \frac{{D}}{{\delta^k  \Delta}}.\label{Assume_R}
\end{equation}
With other concurrent links' transmission power equal to $P_{max}$, the transmission power needed for link $(s_u,u)$ to achieve  $R_{{s_u}u}''$ can be obtained as

\begin{equation}
\begin{array}{l}
P_t^u = \hspace{-0.1cm}\frac{{\left({2^{\frac{{R_{{s_u}u}''}}{{\eta W}}}} \hspace{-0.1cm}- 1\right)\left({{N_0}W + \rho \hspace{-0.4cm}\sum\limits_{v \in {V^k{\backslash \{ u\}}}} \hspace{-0.4cm}{ k_0{G_t}(s_v,u){G_r}(s_v,u){l_{s_vu}}^{ - \tau }{P_{max}}} } \right)}}{{{ {k_0}{G_t}({s_u},u){G_r}({s_u},u){l_{{s_u}u}}^{ - \tau }}}}. \label{P_final}
\end{array}
\end{equation}

For the example in Fig. \ref{fig:EMS operation}, the power control algorithm adjusts the transmission power of each link, and allocates time slots for each pairing.
For the first pairing, two time slots are allocated. Three time slots are allocated to the second pairing, and three time slots are allocated to the third pairing.
In each pairing, each transmission occupies the time slots, and the transmission power is adjusted to reduce energy consumption.
Since the algorithm performs power control on each link,
it achieves a computational complexity of $\mathcal{O}(|V|)$.


\section{Performance Analysis}\label{S5+}

In this section, we demonstrate the roles of D2D communications and concurrent transmissions in reducing energy consumption via theoretical analysis.

For each user $u$, in the serial unicast scheme, we can obtain the transmission rate from the BS to $u$ as

\begin{equation}
\begin{array}{l}
{R_{\alpha u}} = \eta W{\log _2}\left(1 + \frac{{{k_0}{G_t}(\alpha,u){G_r}(\alpha,u){l_{\alpha u}}^{ - \tau }P_{max}}}{{{N_0}W  }}\right),
\end{array}\label{rate-o}
\end{equation}
where we use $\alpha$ to denote the base station.
Then we can obtain the energy consumption to provide multicast service for user $u$ as
\begin{equation}
E^s_{u}=P_{max} \cdot {\frac{D}{{{R_{\alpha u}}}}} .
\end{equation}

In EMS, in contrast, the transmission power to serve user $u$ is

\begin{equation}
\begin{array}{l}
P_t^u = \hspace{-0.1cm}\frac{{\left({2^{\frac{{R_{{s_u}u}''}}{{\eta W}}}} \hspace{-0.1cm}- 1\right)\left({{N_0}W + \rho \hspace{-0.4cm}\sum\limits_{v \in {V^k{\backslash \{ u\}}}} \hspace{-0.4cm}{ k_0{G_t}(s_v,u){G_r}(s_v,u){l_{s_vu}}^{ - \tau }{P_{max}}} } \right)}}{{{ {k_0}{G_t}({s_u},u){G_r}({s_u},u){l_{{s_u}u}}^{ - \tau }}}}. \label{P_final1}
\end{array}
\end{equation}
With (\ref{Assume_R}) incorporated, we can obtain

\begin{equation}
\begin{array}{l}
P_t^u = \hspace{-0.1cm}\frac{{\left({2^{\frac{{{D}}}{{\delta^k  \Delta\eta W}}}} \hspace{-0.1cm}- 1\right)\left({{N_0}W + \rho \hspace{-0.4cm}\sum\limits_{v \in {V^k{\backslash \{ u\}}}} \hspace{-0.4cm}{ k_0{G_t}(s_v,u){G_r}(s_v,u){l_{s_vu}}^{ - \tau }{P_{max}}} } \right)}}{{{ {k_0}{G_t}({s_u},u){G_r}({s_u},u){l_{{s_u}u}}^{ - \tau }}}}. \label{P_final2}
\end{array}
\end{equation}
Then we can obtain the energy consumption to serve $u$ in EMS as $E^d_{u}=P_t^u \cdot \delta^k \cdot \Delta$.
With (\ref{P_final2}) incorporated, we can obtain
\begin{equation}
\begin{aligned}
&E^d_{u}=\left({2^{\frac{{{D}}}{{\delta^k  \Delta\eta W}}}} \hspace{-0.1cm}- 1\right)\cdot\\&\hspace{-0.1cm}\frac{{\left({{N_0}W + \rho \hspace{-0.4cm}\sum\limits_{v \in {V^k{\backslash \{ u\}}}} \hspace{-0.4cm}{ k_0{G_t}(s_v,u){G_r}(s_v,u){l_{s_vu}}^{ - \tau }{P_{max}}} } \right)}}{{{ {k_0}{G_t}({s_u},u){G_r}({s_u},u){l_{{s_u}u}}^{ - \tau }}}} \cdot \delta^k \cdot \Delta. \label{energy-o}
\end{aligned}
\end{equation}
Since the interference between links is controlled by the interference threshold $\sigma$, we can obtain
\begin{equation}
E^d_{u} < \hspace{-0.1cm}\frac{{\left({2^{\frac{{{D}}}{{\delta^k  \Delta\eta W}}}} \hspace{-0.1cm}- 1\right)\left({{N_0}W + (|V^k|-1)\sigma {P_{max}}} \right)}}{{{ {k_0}{G_t}({s_u},u){G_r}({s_u},u){l_{{s_u}u}}^{ - \tau }}}} \cdot \delta^k \cdot \Delta.   \label{energy}
\end{equation}

The right side of (\ref{energy}) can be regarded as an upper bound of $E^d_{u}$. To minimize the energy consumption, we can try to minimize the upper bound. First, we investigate the role of D2D communications in our scheme. With the antennas between the transmitter and receiver towards each other, the benefits of D2D communications relay on the term ${l_{{s_u}u}}^{ - \tau }$. Thus, where D2D communications are enabled, we usually have $l_{{s_u}u} < l_{\alpha u}$, and the shorter $l_{{s_u}u}$ is, the more benefits we can obtain from D2D communications. Therefore, in the D2D path planning algorithm, we establish the D2D paths by searching the nearest neighbors of last nodes on paths.

Then we analyze the role of concurrent transmissions on our scheme. To minimize the energy consumption, we can try to minimize the right side of (\ref{energy}). In the serial unicast scheme, there is only one link in each pairing, and $|V^k|$ is equal to 1. Through concurrent transmissions, the interference increases as shown in (\ref{energy-o}), but the interference is controlled by the threshold $\sigma$. After concurrent transmission scheduling, more time slots can be distributed to each link, and $\delta^k$ increases. Thus, we can observe concurrent transmissions increase the number of time slots scheduled for each user with the cost of increased interference between links. With the interference between links controlled by the interference threshold, the energy consumption can be reduced by increasing the number of time slots significantly. To simplify the notation, we define $\gamma$ as

\begin{equation}
\gamma=\hspace{-0.1cm}\frac{{\left({{N_0}W + (|V^k|-1)\sigma {P_{max}}} \right)}}{{{ {k_0}{G_t}({s_u},u){G_r}({s_u},u){l_{{s_u}u}}^{ - \tau }}}}  \cdot \Delta.
\end{equation}
Then we can denote the right side of (\ref{energy}) as $E_u$, which can be obtained as
\begin{equation}
E_u=\gamma \cdot \left({2^{\frac{{{D}}}{{\delta^k  \Delta\eta W}}}} \hspace{-0.1cm}- 1\right)\cdot \delta^k.
\end{equation}
Taking the derivative of $E_u$ respect to $\delta^k$, we have
\begin{equation}
\frac{d E_u}{d \delta^k}=\gamma \left(2^{\frac{{{D}}}{{\delta^k  \Delta\eta W}}}\left(1-{\rm{ln}} 2{\frac{{{D}}}{{\delta^k  \Delta\eta W}}}\right)-1\right).
\end{equation}
Denoting ${\frac{{{D}}}{{\delta^k  \Delta\eta W}}}$ by $x$, we know $2^x\left(1-{\rm{ln}}2\cdot x\right)-1$ is a strictly monotone decreasing function, and since $2^x\left(1-{\rm{ln}}2\cdot x\right)-1$ is equal to 0 when $x$ is equal to 0, we know $\frac{d E_u}{d \delta^k}<0$ when ${\frac{{{D}}}{{\delta^k  \Delta\eta W}}}>0$, which always holds in our case. Therefore, we can always reduce the energy consumption by allocating more time slots to each user. After concurrent transmission scheduling, links are grouped into a few number of pairings, and more time slots are allocated to each transmission, which will reduce the energy consumption.
To ensure the multicast throughput not less than that achieved in the serial unicast scheme, we require the total number of time slots cannot exceed that in the serial unicast scheme, and thus we distribute the $T_s$ time slots to all pairings. To ensure the energy consumption for each user is reduced, we distribute the time slots to pairings proportionally to their needed number of time slots as indicated by (\ref{condition}).

For each user $u$, to ensure the energy consumption in EMS is less than that in the serial unicast scheme, we require $E^d_{u}<E^s_{u}$. Denoting the $\rho \hspace{-0.4cm}\sum\limits_{v \in {V^k{\backslash \{ u\}}}} \hspace{-0.4cm}{ k_0{G_t}(s_v,u){G_r}(s_v,u){l_{s_vu}}^{ - \tau }{P_{max}}} $ by $I_u$ to simplify notation, we can obtain the condition for $I_u$ to satisfy $E^d_{u}<E^s_{u}$ as follows.

\begin{equation}
I_u<\frac{DP_{max}{k_0}{G_t}({s_u},u){G_r}({s_u},u){l_{{s_u}u}}^{ - \tau }}{{R_{\alpha u}}\left({2^{\frac{{{D}}}{{\delta^k  \Delta\eta W}}}} \hspace{-0.1cm}- 1\right)\delta^k\Delta}.
\end{equation}
From the equation, we can observe that for shorter D2D links, the tolerant interference for this link is larger. Besides, since $\left({2^{\frac{{{D}}}{{\delta^k  \Delta\eta W}}}} \hspace{-0.1cm}- 1\right)\delta^k$ decreases with $\delta^k$, and when allocating more time slots to this link, the tolerant interference is larger.

After power control, more time slots are allocated to each user, and
the transmission power is reduced, and the achieved SINR is low. If we approximate $\left({2^{\frac{{{D}}}{{\delta^k  \Delta\eta W}}}} \hspace{-0.1cm}- 1\right)$ by ${\frac{{{D}}}{{\delta^k  \Delta\eta W}}}{\rm{ln}}2$, then (\ref{energy-o}) can be expressed as

\begin{equation}
E^d_{u}\approx\hspace{-0.1cm}\frac{{{{D}}}{\rm{ln}}2{\left({{N_0}W + I_u} \right)}}{{{\eta W {k_0}{G_t}({s_u},u){G_r}({s_u},u){l_{{s_u}u}}^{ - \tau }}}} . \label{energy-o-a}
\end{equation}
We can observe that the energy consumption increases linearly with the multicast traffic demand $D$, Which is consistent with our performance evaluation results in Fig. \ref{fig:EC-D}.


\section{Performance Evaluation}\label{S6}

In this section, we evaluate the performance of EMS and compare it against other schemes under various system parameters. Besides, the impact of the threshold on the throughput of EMS is also investigated.

\subsection{Simulation Setup}\label{S6-1}

In a typical mmWave small cell, we assume the BS is located in the center of a square area of $20 m \times 20 m$, where several users are uniformly distributed. In the simulation, we adopt the reference antenna model with side lobe in IEEE 802.15.3c, which consists of
a main lobe of the Gaussian form and constant level of side lobes \cite{chen_2}. The antenna gain in decibels (dBs), $G(\theta )$, can be expressed as
\begin{equation}
G(\theta ) = \left\{ {\begin{array}{*{20}{c}}
{{G_0} - 3.01 \cdot {{(\frac{{2\theta }}{{{\theta _{ - 3{\rm{dB}}}}}})}^2},\;0^ \circ \le \theta  \le {\theta _{ml}}/2};\\
{{G_{sl}},\hspace{2.5cm}{\theta _{ml}}/2 \le \theta  \le {{180}^ \circ }},
\end{array}} \right.
\end{equation}
where $\theta$ is an arbitrary angle within the range $[0^ \circ, 180 ^ \circ]$. ${{\theta _{ - 3{\rm{dB}}}}}$ is the angle of the half-power beamwidth, and ${{\theta _{ml}}}$ is the main lobe width in units of degrees. ${{\theta _{ml}}}$ is related to ${{\theta _{ - 3{\rm{dB}}}}}$ via ${\theta _{ml}} = 2.6 \cdot {\theta _{ - 3{\rm{dB}}}}$.
The maximum antenna gain ${{G_0}}$ can be obtained from ${{\theta _{ - 3{\rm{dB}}}}}$ as ${G_0} = 10\log {(\frac{{1.6162}}{{\sin ({\theta _{ - 3{\rm{dB}}}}/2)}})^2}$, while the side lobe gain ${{G_{sl}}}$ can be obtained by ${G_{sl}} =  - 0.4111 \cdot \ln ({\theta _{ - 3{\rm{dB}}}}) - {\rm{10}}{\rm{.579}}$. Of course, there are differences between theoretical directional antenna model and practical phased-array beam patterns, and
strong side lobes of consumer-grade antennas may weaken the advantages of concurrent transmissions in our scheme \cite{practicalAntenna}.

We summarize the simulation parameters in Table \ref{tab:para-EMS}. For every result, we perform 50 independent experiments, and the mean of the results are plotted in the figures.

\begin{table}
\begin{center}
\caption{Simulation Parameters}
\def \temptablewidth {0.9\textwidth}
\begin{tabular}{ccc}
\hline
\textbf{Parameter}&\textbf{Symbol}&\textbf{Value}\\
\hline
Maximum transmission power & $P_{max}$ & 30 dBm\\
Bandwidth & W & 2160 MHz \\
Noise power spectra density &$N_0$& -134dBm/MHz\\
Path loss exponent & $\tau $ & 2\\
Time slot duration &$\Delta $& 18 $\mu$s\\
MUI factor & $\rho$ &  1\\
Half-power beamwidth & ${{\theta _{ - 3{\rm{dB}}}}}$ & ${\rm{15}}^\circ $\\
Efficiency of the transceiver design &$\eta$ & 0.5\\
Interference Threshold & $\sigma$ & ${10^{ - 12}}$ \\
Maximum number of hops & $H_m$ & 6\\
Multicast data size & $D$ & $10^9$ bit\\
Multicast group size &$|\mathbb{U}|$ & 15\\
\hline
\end{tabular}
\label{tab:para-EMS}
\end{center}
\end{table}

In the evaluation, our scheme EMS is compared with the following two multicast schemes:

1) \emph{\textbf{FDMAC}}: the frame-based scheduling directional MAC protocol \cite{mao}. In FDMAC, since D2D communications are not considered, and links from the BS to users are adjacent, FDMAC will be reduced to the serial unicast scheme. From the comparison with FDMAC, we can observe the advantages of concurrent transmissions and D2D communications in our proposed scheme.



2) \emph{\textbf{D2D}}: the D2D multicast scheme, where D2D communications are exploited as EMS to improve system performance. In the D2D scheme, the D2D paths are selected the same as EMS, but the concurrent transmissions are not enabled in the D2D scheme. After selecting the D2D paths,
the links on the D2D paths are scheduled into each pairing in sequence (the inherent transmission order of links on each D2D path should be ensured), and there is one link in each pairing. Then the transmission power of each link is adjusted in the same way as EMS with the only difference of only one link in each pairing. From the comparison with the D2D multicast scheme, we can observe the role concurrent transmissions play in our scheme.

In the evaluation, we consider three performance metrics as follows.

%

1) \textbf{Energy Consumption:} Total energy consumption of multicast transmissions in the network, which is denoted by $EC$, and
can be expressed as

\begin{equation}
EC = \sum\limits_{k = 1}^K {\sum\limits_{u\in V^k}{P_t^u  \delta^k  \Delta}}.
\end{equation}

2) \textbf{Energy Ratio:} Energy consumption of EMS divided by that of the D2D Scheme. We denote energy ratio by $ER$, which can be expressed as

\begin{equation}
ER = \frac{{E{C_{EMS}}}}{{E{C_{D2D}}}},
\end{equation}
where we denote the energy consumption of EMS and D2D scheme by ${E{C_{EMS}}}$ and ${E{C_{D2D}}}$, respectively.


3) \textbf{D2D Ratio:} Energy consumption using D2D communications divided by the total energy consumption for multicast service.

In this paper, we define the energy efficiency as the achieved throughput divided by the consumed energy. With the throughput ensured, the energy efficiency is mainly determined by the energy consumption since in our scheme the time slots consumed by the serial unicast scheme is distributed to the multicast service for users as indicated by (\ref{condition}).

\subsection{Time and Energy Overhead in Beam Training} \label{S6-2}

For the beam training in the scheduling part of EMS, we assume the beam training is done with the help of location information from location techniques \cite{MoeWin-location,mmW-V2V}.
Compared with the serial unicast scheme, EMS and the D2D scheme need to perform beam training between D2D pairs after the D2D paths are established. When the beam training for D2D pairs on the path is completed, the BS computes a schedule by the MIS based multihop scheduling algorithm and the power control algorithm, and then pushes the schedule to users in the multicast group.


With the location information of nodes, the average number of training beam pairs can be significantly reduced, and we adopt 10 candidate beam pairs in the simulation
\cite{mmW-V2V}. For each D2D pair, the BS first transmits a small control packet to inform the transmitter of the candidate beam pairs and the corresponding receiver.
Then the BS also transmits a small control packet to inform the receiver of the candidate beam pairs and the corresponding transmitter.
Afterwards,
the transmitter and receiver transmit one small control packet using each training beam pair, and the receiver records the received SNR and transmits an acknowledgement packet with the recorded SNR to the transmitter. After all the candidate training beam pairs are used to perform beam training, the beam pair with the highest received SNR is adopted for the transmission between the transmitter and the receiver. We adopt the simulation parameters in \cite{MRDMAC}, and beam training using one beam pair can be completed within $T_{ShFr}+T_{SIFS}+T_{ACK}$. The transmission power in beam training is 30 dBm.
Detailed parameters for beam training is listed in Table \ref{tab:para-BT}.
For EMS and the D2D scheme, additional beam training between D2D pairs is needed to execute D2D communications in the transmission part.


\begin{table}
\begin{center}
\caption{Beam Training Parameters}
\def \temptablewidth {0.9\textwidth}
\begin{tabular}{ccc}
\hline
\textbf{Parameter}&\textbf{Symbol}&\textbf{Value}\\
\hline
Transmission rate in beam training & $R$ & 2 Gbps \\
Propagation delay&${\delta _p}$& 50ns\\
PHY overhead& ${T_{PHY}}$ & 250ns\\
Short MAC frame Tx time& ${T_{ShFr}}$& ${T_{PHY}}$+$14*8/R$+${\delta _p}$\\
SIFS interval&${T_{SIFS}}$& 100ns\\
ACK Tx time&${T_{ACK}}$&${T_{ShFr}}$\\
\hline
\end{tabular}
\label{tab:para-BT}
\end{center}
\end{table}

\begin{figure}[htbp]
\begin{minipage}[t]{1\linewidth}
\centering
\includegraphics[width=0.85\columnwidth]{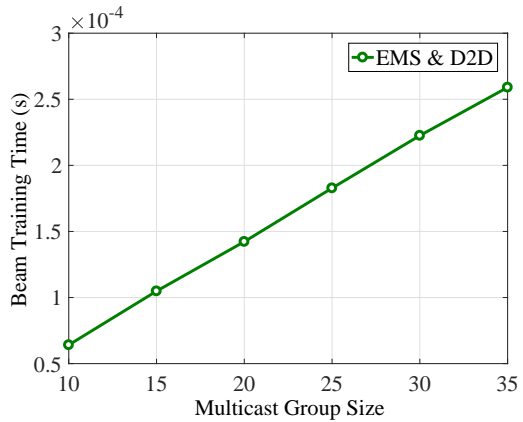}
\end{minipage}%
\caption{The time overhead for beam training between D2D pairs in EMS and the D2D scheme under different multicast group sizes.}
\label{fig:TO-B} 
\vspace*{-3mm}
\end{figure}

In Fig. \ref{fig:TO-B}, we plot the additional time overhead for beam training between D2D pairs in EMS and the D2D scheme under different multicast group sizes. With the increase of multicast group size, the time overhead increases due to more D2D pairs on the D2D paths. When the multicast group size is 35, the time overhead for beam training of D2D pairs is about $2.59 \times 10^{-4}$ s, while the duration of the transmission part is about 7.94 s.
We can also observe that the time overhead is significantly smaller than the duration of the transmission part. Therefore, the additional time overhead for beam training of D2D pairs has a marginal impact on the overall throughput.

\begin{figure}[htbp]
\begin{minipage}[t]{1\linewidth}
\centering
\includegraphics[width=0.85\columnwidth]{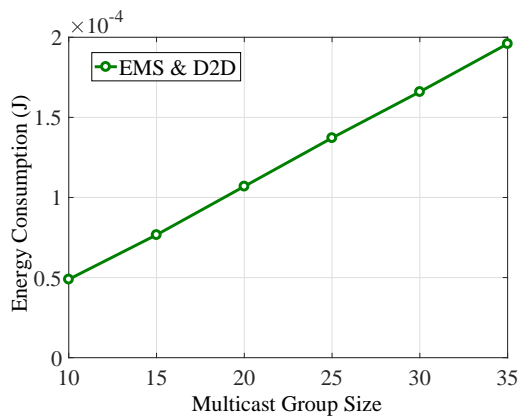}
\end{minipage}%
\caption{The energy consumption for beam training between D2D pairs in EMS and the D2D scheme under different multicast group sizes.}
\label{fig:EC-B} 
\vspace*{-3mm}
\end{figure}

We also plot the additional energy consumption for beam training between D2D pairs in EMS and the D2D scheme under different multicast group sizes in Fig. \ref{fig:EC-B}. The energy consumption also increases with the multicast group size due to more D2D pairs on D2D paths.
When the multicast group size is 35, the energy consumption is about $2 \times 10^{-4}$ J, and the energy consumption in the transmission part is about 1.2243 J, which is shown in Fig. \ref{fig:EC-UN}. Thus, there is a minor increase in the energy consumption due to beam training of D2D pairs.
Therefore, combining the results in Fig. \ref{fig:TO-B}, beam training of D2D pairs in EMS and the D2D scheme has a marginal impact on the overall throughput, energy consumption, and the energy efficiency. In the following, we focus on the energy consumption, throughput, and energy efficiency in the transmission part.

\subsection{Energy Consumption Comparison in the Transmission Part} \label{S6-2}

\begin{figure}[htbp]
\begin{minipage}[t]{1\linewidth}
\centering
\includegraphics[width=0.85\columnwidth]{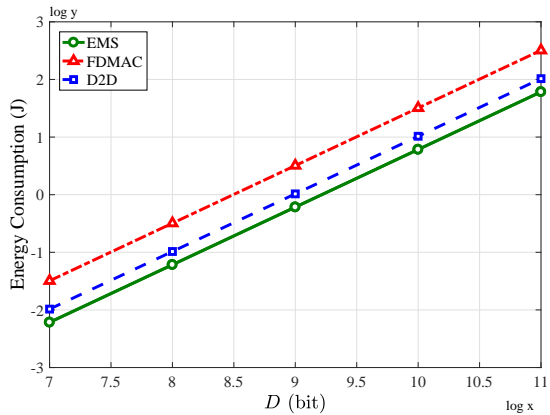}
\end{minipage}%
\caption{The energy consumption comparison of three schemes under different multicast data sizes.}
\label{fig:EC-D} 
\vspace*{-3mm}
\end{figure}

The energy consumption of three schemes under different multicast data sizes is plotted in Fig. \ref{fig:EC-D}. To show the gap between different schemes more clearly, we show the results with logarithmic coordinates. Other parameters are the same as Table \ref{tab:para-EMS} except the multicast data size. We can observe that EMS achieves the lowest energy consumption, and the gap between the D2D scheme and FDMAC shows the role of D2D communications in reducing the energy consumption. Considering the three schemes achieve the same throughput in the transmission part since the duration of the transmission part is the same, the energy efficiency of EMS is the highest due to the lowest energy consumption.
At the same time, the gap between the D2D scheme and EMS shows the role of concurrent transmissions in reducing the energy consumption. We can also observe that the energy consumption increases with the increase of multicast data size. When the multicast data size is larger, more transmission time is needed to complete the multicast task, and thus the energy consumption increases. Compared with the D2D scheme, EMS reduces the energy consumption by about 41.1\% when the multicast data size is $10^{11}$ bit. When the multicast data size is $10^{11}$ bit, EMS reduces the energy consumption by about 81\% compared with FDMAC.

\begin{figure}[htbp]
\begin{minipage}[t]{1\linewidth}
\centering
\includegraphics[width=0.85\columnwidth]{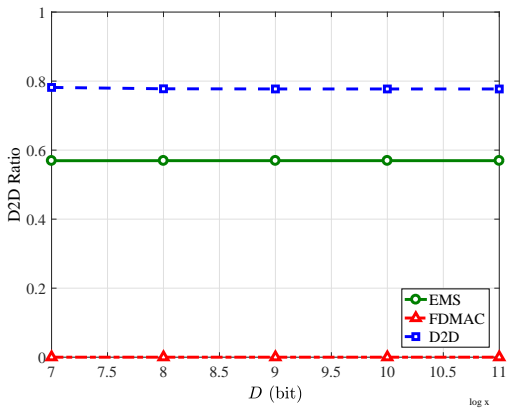}
\end{minipage}%
\caption{The D2D ratio comparison of three schemes under different multicast data sizes.}
\label{fig:DR-D} 
\vspace*{-3mm}
\end{figure}

The D2D ratio comparison of three schemes under different multicast data sizes is plotted in Fig. \ref{fig:DR-D}. From the results, we can observe
that the D2D scheme achieves the highest D2D ratio among the schemes. Since D2D communications are not enabled in FDMAC, its D2D ratio is 0. In contrast, our proposed EMS achieves a relatively low D2D ratio. In EMS, most multicast transmissions are via D2D communications, and
due to the concurrent transmissions, energy consumption using D2D communications is reduced much lower than the D2D scheme without concurrent transmissions. Thus, EMS achieves lower D2D ratio than the D2D scheme. For each transmission, the energy consumption is proportional to the multicast data size, and thus the D2D ratio remains constant with the multicast data size.

\begin{figure}[htbp]
\begin{minipage}[t]{1\linewidth}
\centering
\includegraphics[width=0.85\columnwidth]{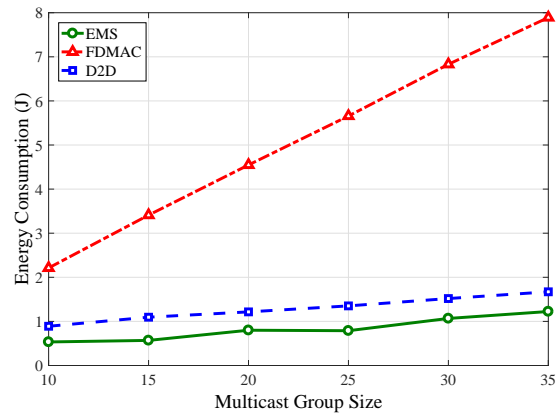}
\end{minipage}%
\caption{The energy consumption comparison under different multicast group sizes.}
\label{fig:EC-UN} 
\vspace*{-3mm}
\end{figure}

The energy consumption comparison of three schemes under different multicast group sizes is plotted in Fig. \ref{fig:EC-UN}. With the increase of the multicast group size, the energy consumption of three schemes increases due to more users should be served. However, the energy consumption of EMS and the D2D scheme increases with the multicast group size slowly, which demonstrates the advantages of D2D communications and concurrent transmissions in our scheme.
 As the multicast group size increases, the density of devices increases, and more interference can be observed. Since the contention graph is constructed based on the interference between links, fewer concurrent transmissions are allowed. On the other hand, with the increase of density of devices, there are better channels between devices, and the advantages of D2D communications become larger. Therefore, concurrent transmissions and D2D communications affect the energy consumption in an opposite way, and the energy consumption has a slow rising tendency with the increase of the multicast group size, which indicates a bigger role played by fewer concurrent transmissions and more users.
 As we can observe, the D2D scheme already improves the performance to a large extent, and when the multicast group size is 35, the D2D scheme reduces the energy consumption by about 78.8\% compared with FDMAC, which demonstrates obvious advantage of exploiting D2D communication in improving energy efficiency. Compared with the D2D scheme, EMS further reduces the energy consumption by about 27\% due to the concurrent transmission mechanism in EMS. From the results, we can observe that D2D communications play a bigger role in reducing energy consumption than concurrent transmissions for EMS.

\begin{figure}[htbp]
\begin{minipage}[t]{1\linewidth}
\centering
\includegraphics[width=0.85\columnwidth]{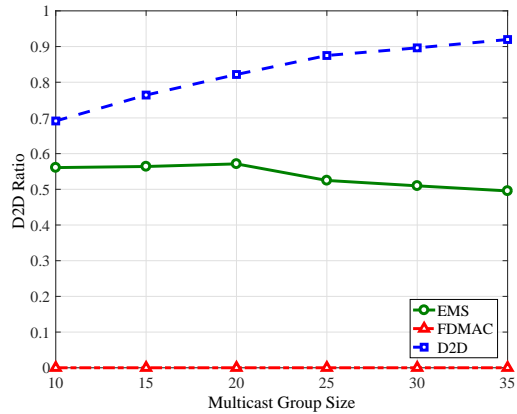}
\end{minipage}%
\caption{The D2D ratio comparison under different multicast group sizes.}
\label{fig:DR-UN} 
\vspace*{-3mm}
\end{figure}

The D2D ratio comparison under different multicast group sizes is plotted in Fig. \ref{fig:DR-UN}. Consistent with the results in Fig. \ref{fig:DR-D}, EMS achieves a relatively low value due to the enabled concurrent transmissions. The tendency for EMS and the D2D scheme is different. With more users to serve, there are more multicast services are via D2D communications, and thus the D2D ratio for the D2D scheme increases.
For EMS, more users lead to more concurrent transmissions, and the energy consumption using D2D communications can be reduced to a larger extent. Thus, the achieved D2D ratio decreases slowly with the multicast group size.

\begin{figure}[htbp]
\begin{minipage}[t]{1\linewidth}
\centering
\includegraphics[width=0.85\columnwidth]{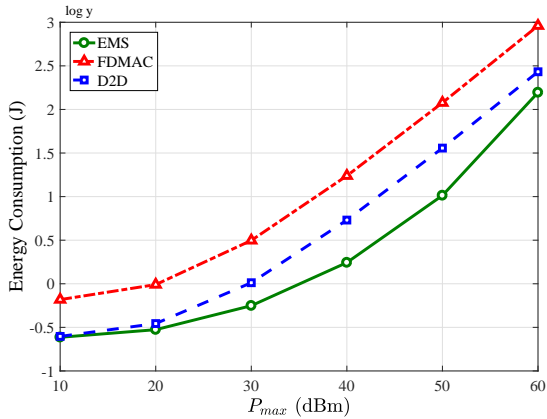}
\end{minipage}%
\caption{The energy consumption comparison under different maximum transmission power.}
\label{fig:EC-Pt} 
\vspace*{-3mm}
\end{figure}

In Fig. \ref{fig:EC-Pt}, we plot the energy consumption comparison under different maximum transmission power. The results are shown with Y-axis using the logarithmic coordinates. With the increase of the maximum transmission power, the energy consumption of three schemes increases since the number of time slots occupied by FDMAC, $T_s$, decreases due to higher transmission rates. With lower
$T_s$, fewer time slots can be allocated to each transmission in EMS and the D2D scheme. Thus, much reduction in the transmission power cannot be achieved, and the power control mechanism in EMS and the D2D scheme cannot play a big role in reducing the energy consumption. Therefore, the energy consumption increases with the maximum transmission power. Although lower maximum transmission power reduces the energy consumption, it also leads to lower network throughput. Thus, the maximum transmission power should be selected according to practical throughput and energy consumption requirements. As we can observe, EMS still has the lowest energy consumption, and the highest energy efficiency.


\begin{figure}[htbp]
\begin{minipage}[t]{1\linewidth}
\centering
\includegraphics[width=0.85\columnwidth]{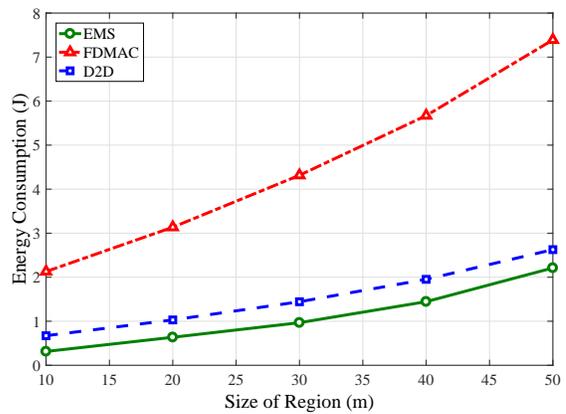}
\end{minipage}%
\caption{The energy consumption comparison under different region sizes.}
\label{fig:EC-Size} 
\vspace*{-3mm}
\end{figure}

In Fig. \ref{fig:EC-Size}, we plot the energy consumption comparison under different region sizes, where X-axis represents the side length of the square region in the unit of meter (m). EMS achieves the lowest energy consumption among the three schemes under different region sizes. With larger region size, the energy consumption increases since users are distributed more dispersedly, and the link length also increases, which increases the propagation loss. With higher propagation loss, more energy is needed to complete the multicast service. When users are distributed more dispersedly, less interference exists between links, and more concurrent transmissions can be enabled. However, more energy consumption from larger link length plays a dominated role considering the high propagation loss at mmWave bands, and the energy consumption increases with the size of region.
When the size of region is 50m, EMS reduces the energy consumption by about 70.1\% compared with FDMAC, and by about
16\% compared with the D2D scheme. From the results, we also can observe that the advantage of EMS compared with FDMAC is mainly because of the D2D communication mechanism since the big performance improvement achieved by the D2D scheme.

\begin{figure}[htbp]
\begin{minipage}[t]{1\linewidth}
\centering
\includegraphics[width=0.85\columnwidth]{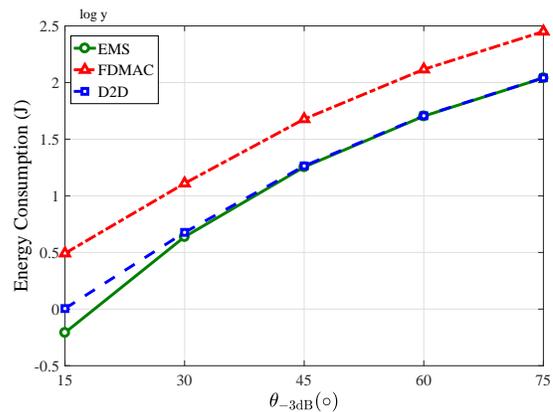}
\end{minipage}%
\caption{The energy consumption comparison under different $\theta _{ - 3{\rm{dB}}}$.}
\label{fig:EC-theta} 
\vspace*{-3mm}
\end{figure}

The energy consumption comparison under different $\theta _{ - 3{\rm{dB}}}$ is plotted in Fig. \ref{fig:EC-theta}. We examine five cases, with $\theta _{ - 3{\rm{dB}}}$ equal to ${\rm{15}}^\circ $, ${\rm{30}}^\circ $, ${\rm{45}}^\circ $, ${\rm{60}}^\circ $, and ${\rm{75}}^\circ $. The results are shown with Y-axis using the logarithmic coordinates.
We can observe that the energy consumption increases with the $\theta _{ - 3{\rm{dB}}}$. With larger $\theta _{ - 3{\rm{dB}}}$, lower antenna gain can be achieved to compensate the propagation loss as indicated by the antenna model, and more energy is needed to complete the multicast service.
As before, EMS achieves the lowest energy consumption.
The gap between EMS and the D2D scheme is larger when $\theta _{ - 3{\rm{dB}}}$ is smaller. Narrow antenna beams have higher directivity and lead to less interference between links, which is beneficial for concurrent transmissions in EMS. We also observe the big role of D2D communications in reducing energy consumption.

\begin{figure}[htbp]
\begin{minipage}[t]{1\linewidth}
\centering
\includegraphics[width=0.85\columnwidth]{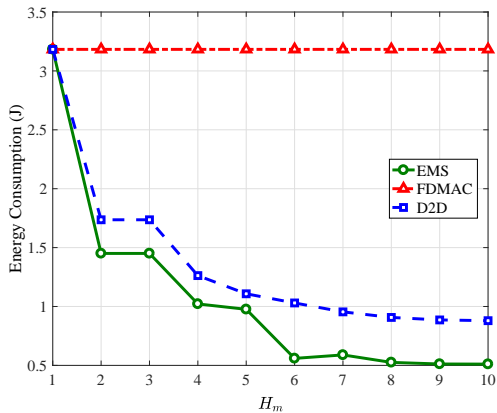}
\end{minipage}%
\caption{The energy consumption comparison under different maximum number of hops.}
\label{fig:EC-Hm} 
\vspace*{-3mm}
\end{figure}

Fig. \ref{fig:EC-Hm} presents the energy consumption comparison under different maximum numbers of hops, $H_m$. With the increase of $H_m$, the energy consumption of EMS and D2D decreases. With a larger maximum number of hops, more users can be served via D2D communications, and more energy can be saved from better channel conditions provided by D2D communications. We also observe that
when $H_m$ is 1, three schemes achieve the same performance since both EMS and D2D reduce to the unicast scheme. When $H_m$ is 1, the concurrent transmission mechanism in EMS is also disabled due to the half-duplex constraint. When $H_m$ increases from 1 to 2, we observe a big decrease of energy consumption for EMS and D2D. When $H_m$ increases to 6, the energy consumption decreases slowly. Since more hops may lead to higher overhead in establishing D2D paths, we select $H_m$ to be 6 to obtain most benefits from D2D communications. Since D2D communications are not enabled in FDMAC, its energy consumption does not change with $H_m$.

Summarizing the results above, EMS reduces energy consumption, and thus improves energy efficiency via D2D communications and concurrent transmissions, and D2D communications play a significant role in reducing energy consumption.



\subsection{Choice of the Interference Threshold} \label{S6-3}

\begin{figure}[htbp]
\begin{minipage}[t]{1\linewidth}
\centering
\includegraphics[width=0.85\columnwidth]{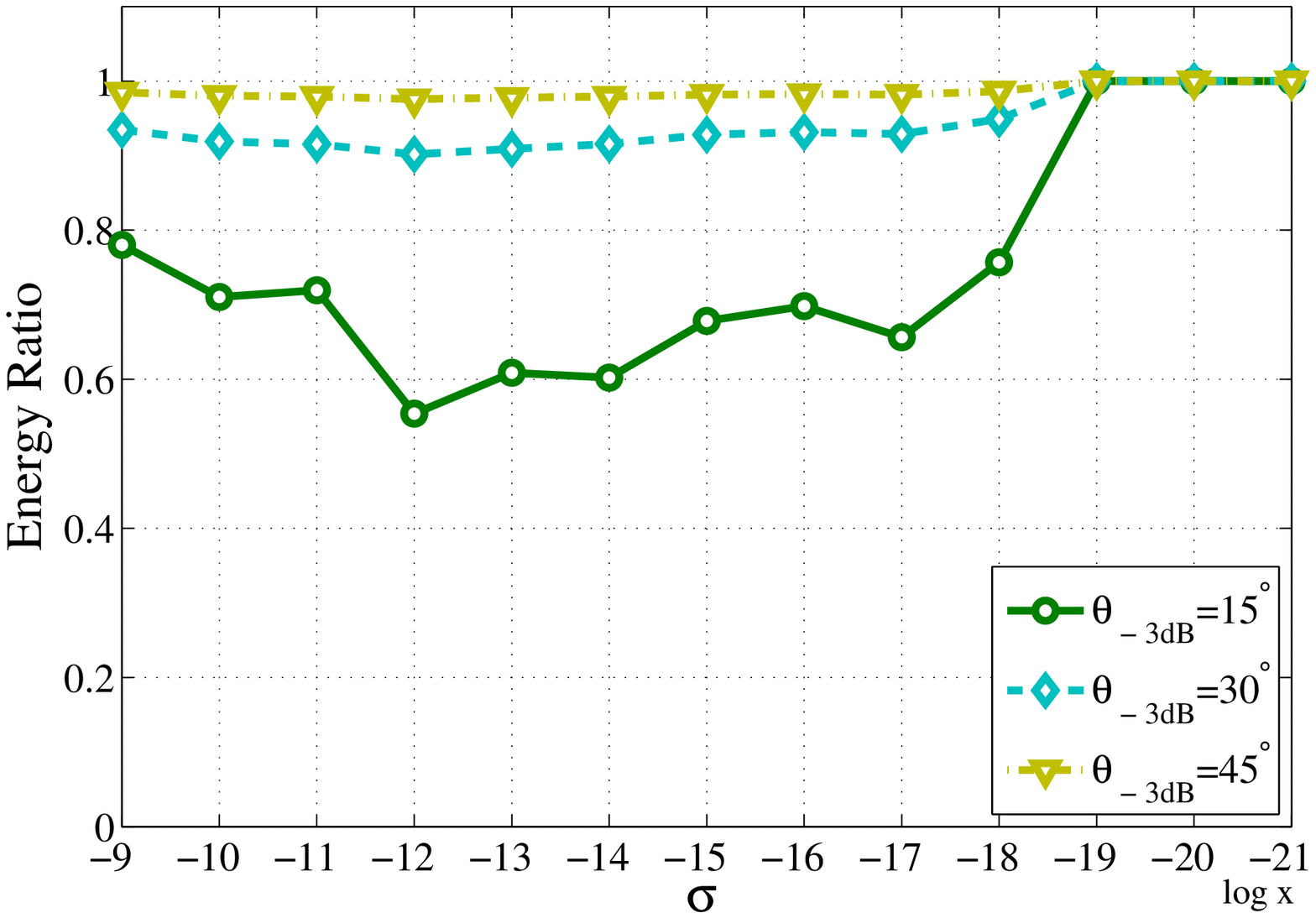}
\end{minipage}%
\caption{The energy ratio performance of EMS with different $\theta _{ - 3{\rm{dB}}}$ under different interference thresholds.}
\label{fig:threshold-theta} 
\vspace*{-3mm}
\end{figure}


Since the choice of threshold affects the concurrent transmission mechanism in EMS, which is the difference between EMS and the D2D scheme,
the energy ratio under different system parameters should be investigated. Fig. \ref{fig:threshold-theta} presents the energy ratio of EMS with different $\theta _{ - 3{\rm{dB}}}$ under different interference thresholds. The results are shown with X-axis using the logarithmic coordinates.
Other parameters are given in Table \ref{tab:para-EMS} except $\theta _{ - 3{\rm{dB}}}$.
We can observe that EMS achieves different lowest energy ratio under different $\theta _{ - 3{\rm{dB}}}$. When $\theta _{ - 3{\rm{dB}}}$ is smaller, the achieved lowest energy ratio can be lower. If the interference
threshold is very small as ${10^{ - 19}}$, the concurrent transmission mechanism is disabled, and EMS reduces to the D2D scheme. Thus, the energy ratio becomes 1. When the threshold is ${10^{ - 12}}$, the energy ratio achieves the almost lowest value in three cases, and therefore we select $\sigma$ to be ${10^{ - 12}}$ in the comparison above.

\begin{figure}[htbp]
\begin{minipage}[t]{1\linewidth}
\centering
\includegraphics[width=0.85\columnwidth]{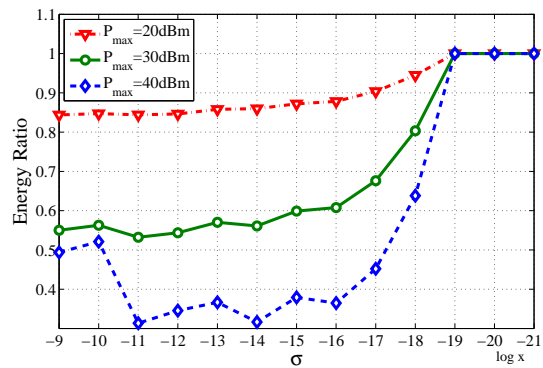}
\end{minipage}%
\caption{The energy ratio performance of EMS with different $P_{max}$ under different interference thresholds.}
\label{fig:threshold-Pt} 
\vspace*{-3mm}
\end{figure}

Fig. \ref{fig:threshold-Pt} gives the energy ratio of EMS with different $P_{max}$ under different interference thresholds.
Other parameters are given in Table \ref{tab:para-EMS} except $P_{max}$. With the increase of $P_{max}$, the achieved lowest energy ratio can be lower. The optimal threshold selection is different for different $P_{max}$. Thus, the threshold should be optimized for different network settings.

\begin{figure}[htbp]
\begin{minipage}[t]{1\linewidth}
\centering
\includegraphics[width=0.85\columnwidth]{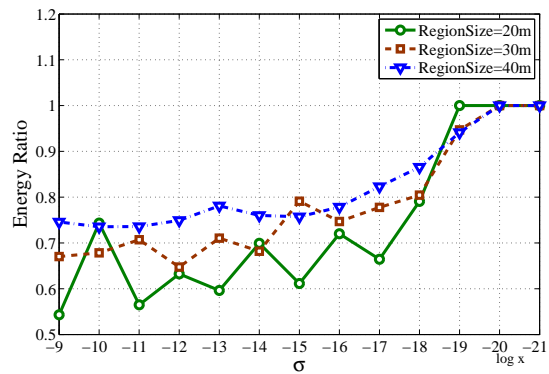}
\end{minipage}%
\caption{The energy ratio performance of EMS with different region sizes under different interference thresholds.}
\label{fig:threshold-RS} 
\vspace*{-3mm}
\end{figure}

The energy ratio of EMS with different region sizes under different interference thresholds is plotted in Fig. \ref{fig:threshold-RS}.
With the decrease of the region size, the achieved lowest energy ratio is lower. At the same time, the optimized thresholds for different region sizes are also different. Generally speaking,
with the decrease of the region size, the interference power relative to $P_{max}$ increases due to less propagation loss, and the optimal threshold also increases.

\section{Conclusions}\label{S7} 

In this paper, we proposed EMS for energy efficient multicast scheduling in mmWave small cells, which exploits both D2D communications and concurrent transmissions to reduce energy consumption. EMS establishes multi-hop D2D transmission paths by the D2D path planning algorithm.
The MIS based concurrent scheduling algorithm schedules the links on the D2D paths into different pairings. Due to more time slots are allocated to each link, lower transmission power can be achieved, and total energy consumption is reduced accordingly. Performance evaluation demonstrates EMS achieves highest energy efficiency compared with other schemes.

Considering the differences between theoretical directional antenna model and practical phased-array beam patterns, we will evaluate the performance of our scheme on a test bed using practical phased-array antennas in the future. Since beamforming in our system relies on the locations of users, we will also investigate the energy consumption involved in finding out pairwise device-device locations. Furthermore, we will also investigate using the multi-level codebook to improve network performance.


 \begin{appendices}
      \section{Notation in Problem Formulation}
      In order to facilitate the reader to understand the notations in Problem Formulation, we list the notations in Table \ref{tab:notation1} as follows.

\begin{table}[htbp]
\begin{center}
\caption{Notation in Problem Formulation}
\def \temptablewidth {0.9\textwidth}
\begin{tabular}{ll}
\hline
\textbf{Symbol}&\textbf{Description}\\
\hline
 $D$ & The traffic demand for the multicast group\\
 $\mathbb{U}$ & The set of users in the multicast group\\
 $u$ & One user in the multicast group\\
 $s_u$ & The transmit node that serves $u$\\
 $\delta ^k$ & The number of time slots for the $k$th pairing\\
 $\Delta$ & The duration of one time slot\\
\multirow{2}{*}{$a_u^k$} & A binary variable to indicate whether the multicast \\&transmission for user $u$ is scheduled in the $k$th pairing\\
$P_t^u$ & The transmission power of $s_u$ to $u$\\
${R_{{s_u}u}^k}$ & The transmission rate for user $u$ in the $k$th pairing\\
\multirow{2}{*}{$T_s$} & The number of time slots needed for \\&the serial unicast scheme to complete the multicast service\\
\hline
\end{tabular}
\label{tab:notation1}
\end{center}
\end{table}

      \section{Notation in EMS}
      To facilitate the understanding of EMS, we also list the notations in Table \ref{tab:notation2} as follows.

\begin{table}[htbp]
\begin{center}
\caption{Notation in EMS}
\def \temptablewidth {0.9\textwidth}
\begin{tabular}{ll}
\hline
\textbf{Symbol}&\textbf{Description}\\
\hline
 $A$ & The base station \\
 $\mathbb{P}$ & The set of selected D2D paths\\
 $p$ & A path in $\mathbb{P}$\\
 $H_m$ & The maximum hop number for each path $p\in\mathbb{P}$\\
 $L_p$ & The last node on $p$\\
 $\mathbb{P}_L$ & The set of last nodes of paths in $\mathbb{P}$\\
 $\sigma$ & The interference threshold\\
 $G_t(V_t,E_t)$ & The contention graph in the $t$th pairing\\
 $V_t$ & The set of vertices\\
 $E_t$ & The set of edges\\
 $d(v)$ & The degree for any vertex $v \in V_t$\\
 $N(v)$ & The set of neighboring vertices for any vertex $v \in V_t$\\
 $V_u^t$ & The set of unvisited and candidate links for the $t$th pairing\\
 \multirow{2}{*}{$T^k_u$} & The number of time slots needed for link \\&$(s_u,u)$ to complete the multicast transmission\\
 \multirow{2}{*}{${T^k}$} & The maximum needed number of time \\&slots for links in the $k$th pairing\\
 \multirow{2}{*}{$R_{{s_u}u}''$} & The needed transmission rate for link\\& $(s_u,u)$ to complete multicast transmission\\
\hline
\end{tabular}
\label{tab:notation2}
\end{center}
\end{table}

  \end{appendices}

\bibliographystyle{IEEEtran}
\end{document}